\documentclass[aps,prl,twocolumn,groupedaddress]{revtex4-1}
\usepackage{graphicx}
\usepackage{amsmath}
\graphicspath{{./Figs/}}

\begin{document}
\title{Barkhausen noise from precessional domain wall motion}
\author{Touko Herranen$^1$}
\author{Lasse Laurson$^2$}
\email{lasse.laurson@tuni.fi}
\affiliation{$^1$Helsinki Institute of Physics, Department of Applied Physics, 
Aalto University, P.O.Box 11100, FI-00076 Aalto, Espoo, Finland.}
\affiliation{$^2$Computational Physics Laboratory, Tampere University, 
P.O. Box 692, FI-33014 Tampere, Finland}

\begin{abstract}
The jerky dynamics of domain walls driven by applied magnetic fields
in disordered ferromagnets -- the Barkhausen effect -- is a
paradigmatic example of crackling noise. 
We study Barkhausen noise in disordered Pt/Co/Pt thin films due to 
precessional motion of domain walls using full micromagnetic simulations, 
allowing for a detailed description of the domain wall internal structure. 
In this regime the domain walls contain topological defects known as Bloch 
lines which repeatedly nucleate, propagate and annihilate within the
domain wall during the Barkhausen jumps.
%making the standard description of domain walls as elastic 
%lines incomplete. 
%These internal degrees of freedom of the domain wall give rise to a violation 
%of the Middleton ``no-passing'' theorem. 
In addition to bursts of domain 
wall propagation, the in-plane Bloch line dynamics within the domain wall 
exhibits crackling noise, and constitutes the majority
of the overall spin rotation activity. 
%This internal dynamics -- neglected by construction
%in typical models of the Barkhausen effect -- constitutes the majority 
%of the overall spin rotation activity. 
\end{abstract}

\maketitle

%\section{Introduction}

Understanding the bursty crackling noise response of elastic objects in 
random media -- domain walls (DWs) \cite{durin2006barkhausen}, 
cracks \cite{laurson2013evolution}, fluids fronts invading porous 
media \cite{rost2007fluctuations}, et cetera -- to slowly 
varying external forces is one of the main problems of statistical 
physics of materials. An important example is given by the magnetic 
field driven dynamics of DWs 
in disordered ferromagnets, where they respond to a slowly 
changing external magnetic field by exhibiting a sequence of discrete jumps 
with a power-law size distribution \cite{durin2006barkhausen,zapperi1998dynamics}. 
This phenomenon, known as the Barkhausen
effect \cite{barkhausen1919zwei}, has been studied extensively, and a 
fairly well-established picture of the 
%possible scenarios of the bursty DW dynamics, or 
possible universality classes of the avalanche 
dynamics, using the language of critical phenomena, is emerging
\cite{durin2006barkhausen,zapperi1998dynamics}. 
%However, especially in 
%thin film geometry Barkhausen noise remains an active research topic, 
%with partially open questions related to the role of a multitude of 
%possible DW structures -- which depend on film thickness and the various 
%magnetic anisotropy contributions -- on the ensuing jerky DW 
%dynamics \cite{laurson2014universality,bohn2013universal,bohn2014statistical,
%mughal2010effect}.

Magnetic DWs constitute a unique system exhibiting crackling noise since
the driving field may, in addition to pushing the wall forward, excite internal
degrees of freedom within the DW \cite{lecomte2009depinning}. 
This effect is well-known especially in the
nanowire geometry -- important for the proposed spintronics devices
such as the racetrack memory \cite{parkin2008magnetic} -- where the 
onset of precession of the DW magnetization
above a threshold field leads to an abrupt drop in the DW 
propagation velocity (the Walker breakdown \cite{schryer1974motion}), 
and hence to a non-monotonic driving field 
vs DW velocity relation \cite{mougin2007domain}; 
these features are well-captured by the so-called
1$d$ models \cite{thiaville2006domain}.
%, describing the DW in terms of 
%scalar variables such as the DW position 
%and the internal DW magnetization angle: Above a Walker field, the latter starts 
%to precess, and the DW propagation velocity drops. 

In wider strips or thin films, the excitations of the DW internal 
magnetization accompanying the velocity drop cannot be described 
by precession of an individual magnetic moment. Instead, one needs 
to consider the nucleation, propagation and annihilation of topological 
defects known as Bloch lines (BLs) within the DW \cite{herranen2015domain,
herranen2017bloch,malozemoff1979magnetic}. 
%For DWs separating two opposite out-of-plane magnetized 
%domains in a ferromagnetic thin film with strong perpendicular magnetic 
%anisotropy (PMA), the Bloch lines correspond to transition regions 
 %separating the two possible chiralities of the Bloch DW. 
BLs, i.e., transition regions separating different 
chiralities of the DW, have been studied in the context of bubble materials 
already in the 1970's \cite{malozemoff1979magnetic}. Their role 
in the physics of the Barkhausen effect needs to be studied.
The typical models of Barkhausen noise, such
as elastic interfaces in random media \cite{zapperi1998dynamics,
laurson2014universality}, scalar field models \cite{caballero2018magnetic} 
or the random field Ising model (RFIM) \cite{perez2004spanning,mughal2010effect,
spasojevic2011avalanche}, exclude BLs by construction.

Here, we focus on understanding the consequences of the presence of BLs %and their dynamics 
within DWs on the jerky DW motion through a disordered thin 
ferromagnetic film. To this end, we study field-driven DW dynamics 
considering as a test system a 0.5 nm thick Co film within a Pt/Co/Pt 
multilayer \cite{metaxas2007creep} with perpendicular magnetic anisotropy 
(PMA) by micromagnetic simulations, able to fully capture the
DW internal structure. %By considering micromagnetic 
%parameters from Ref. \cite{metaxas2007creep}, we adjust the remaining 
%free parameter, 
By tuning the strength of quenched disorder, we match the DW velocity 
vs applied field curve to the experimental one reported in 
\cite{metaxas2007creep}. This leads to a depinning field well 
above the Walker field of the corresponding 
disorder-free system. Hence, when applying a driving scheme corresponding to 
a quasistatic constant imposed DW velocity, the resulting Barkhausen jumps 
take place within the precessional regime. %, such that Bloch lines are created 
%and annihilated during DW propagation. 
%This has a number of consequences 
%which may be contrasted with models assuming a description of the DWs as 
%elastic interfaces in random media without any internal degrees of freedom: 

We find that in addition to avalanches of DW propagation, also the 
in-plane BL magnetization dynamics within the DW exhibits crackling noise, 
and is responsible for the majority of the overall spin rotation activity 
during the Barkhausen jumps; 
the latter dynamics is not directly observable in typical experiments 
(magneto-optical imaging \cite{kim2003direct}, or inductive recording 
\cite{papanikolaou2011universality}). 
The DW can locally move backwards, so it does not obey the Middleton 
no-passing theorem \cite{middleton1992asymptotic}. Functional renormalization
group calculations \cite{le2002two} crucially depend on this property, but 
we find that in line-like DWs BLs do not change the scaling picture of 
avalanches if one looks at measures related to DW displacement. Remarkably,
simple scaling relations applicable to short-range elastic strings in random
media remain valid in the much more complex scenario we consider here. 

%Moreover, the 
%DW, not being a purely elastic line, 
%does not obey the Middleton ``no-passing'' theorem \cite{middleton1992asymptotic} 
%as it can locally move backwards. This poses challenges to 
%theoretical analysis such as functional renormalization group 
%calculations \cite{le2002two} relying on the Middletonian property of 
%the interface. Nevertheless, our results 
%suggest that the presence of Bloch lines -- when considering 
%measures related to DW displacement -- 
%does not change the scaling behaviour of avalanches from that expected 
%for a short-range elastic string in a random medium.

\begin{figure}[t!]
\includegraphics[trim={0 0.7cm 0 0.7cm},clip,width=.9\columnwidth]{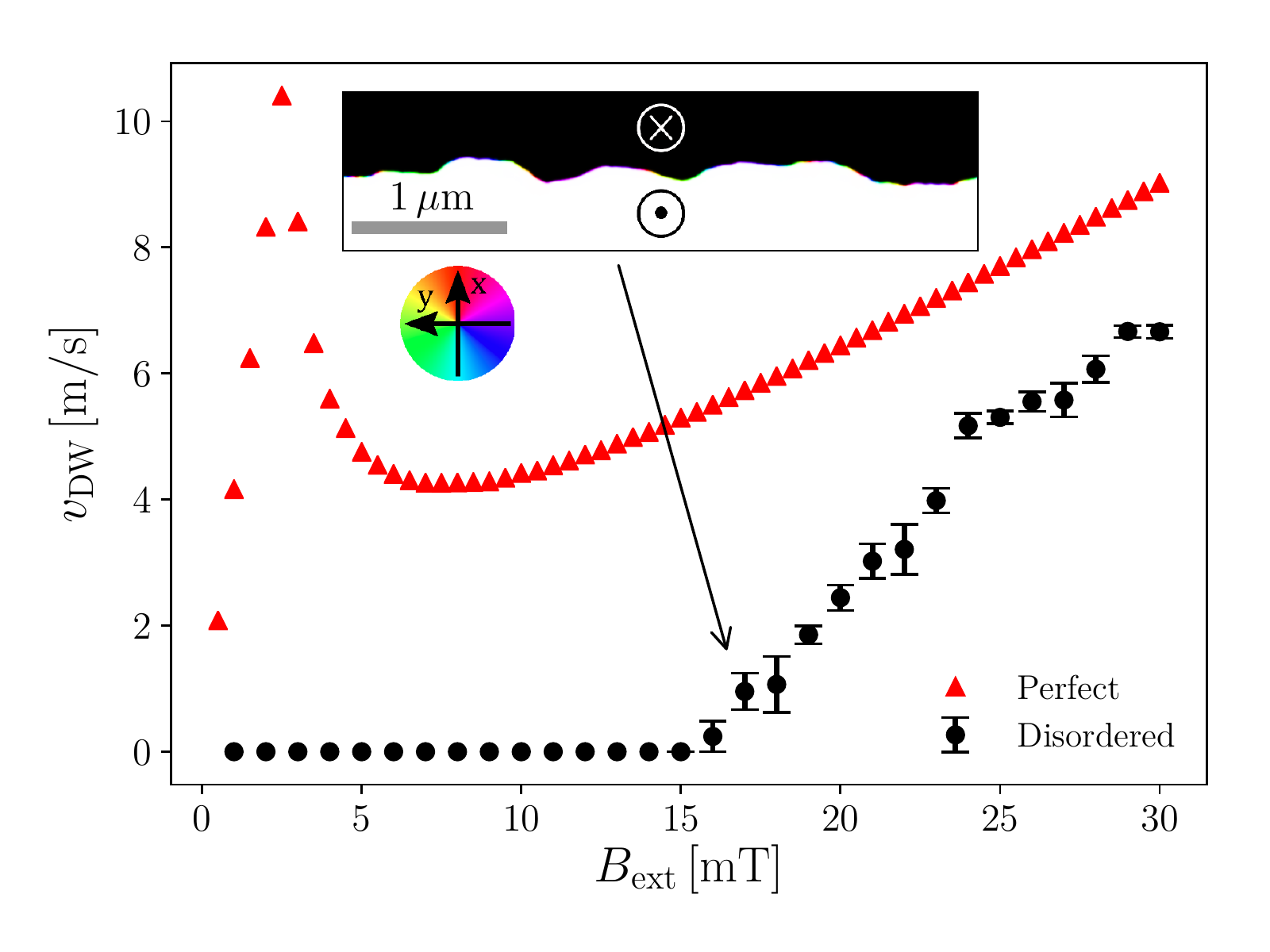}
\caption{$v_\mathrm{DW}$ as a function of $B_\mathrm{ext}$ in a perfect strip 
and in a disordered system where the disorder strength $r$ has been tuned 
to roughly match the $v_\mathrm{DW}(B_\mathrm{ext})$ curve with the
experimental one of Ref. \cite{metaxas2007creep}; the disorder-induced depinning 
field exceeds the Walker field of the perfect strip. Inset shows an example
snapshot of a rough DW containing BLs in the disordered system 
with $B_\text{ext} = 17$ mT.
%Notice that the 
%depinning field due to quenched disorder is significantly larger than the Walker 
%field of the perfect system.
}
\label{fig:scan}
\end{figure}

In our micromagnetic simulations of the DW dynamics, the Landau-Lifshitz-Gilbert
(LLG) equation, $\partial {\bf m}/\partial t =
\gamma {\bf H}_\text{eff} \times {\bf m} + \alpha {\bf m} \times
\partial {\bf m}/\partial t$, describing the time-evolution of the magnetization
${\bf m} = {\bf M}/M_\text{S}$, is solved using the MuMax3 software 
\cite{vansteenkiste2014design}.
In the LLG equation, $\gamma$ is the gyromagnetic ratio, $\alpha$ the Gilbert 
damping parameter, and ${\bf H}_\text{eff}$ the effective field, with 
contributions due to exchange, anisotropy, Zeeman, and demagnetizing energies.
The simulated magnetic material is a 0.5 nm thick Co film in a Pt/Co/Pt 
multilayer with PMA. Micromagnetic parameters for the material are exchange 
stiffness $A_\mathrm{ex} = 1.4\times 10^{-11}\: \mathrm{J/m}$, saturation 
magnetization $M_\mathrm{S} = 9.1 \times 10^5 \: \mathrm{A/m}$, uniaxial 
anisotropy $K_\mathrm{u} = 8.4 \times 10^5 \: \mathrm{J/m^3}$ and damping 
parameter $\alpha = 0.27$; these have been experimentally 
determined in Ref. \cite{metaxas2007creep}.
The resulting DW width parameter is 
$\Delta_\mathrm{DW} = \sqrt{A_\mathrm{ex}/K_0} \approx 7$ nm, 
where $K_0 = K_\mathrm{u} - \frac{1}{2} \mu_0 M_\mathrm{S}^2$ is the 
effective anisotropy. The system size is fixed to $L_x = 1024 \: \mathrm{nm}$, 
$L_y = 4096 \: \mathrm{nm}$ and $L_z = 0.5 \: \mathrm{nm}$. The simulation cell 
dimensions are $\Delta_x = \Delta_y = 2 \: \mathrm{nm}$ and $\Delta_z = 0.5 
\: \mathrm{nm}$. In every simulation the DW, separating domains oriented along
$\pm z$, is initialized along the $+y$ direction as a Bloch wall with the DW 
magnetization in the $+y$ direction. Periodic boundary conditions are used in 
the $y$-direction to avoid boundary effects. The LLG equation 
is then solved using the Dormand-Prince solver (RK45) with an adaptive 
time step.

\begin{figure}[ht!]
\includegraphics[trim={0 0.7cm 0 0},clip,width=.9\columnwidth]{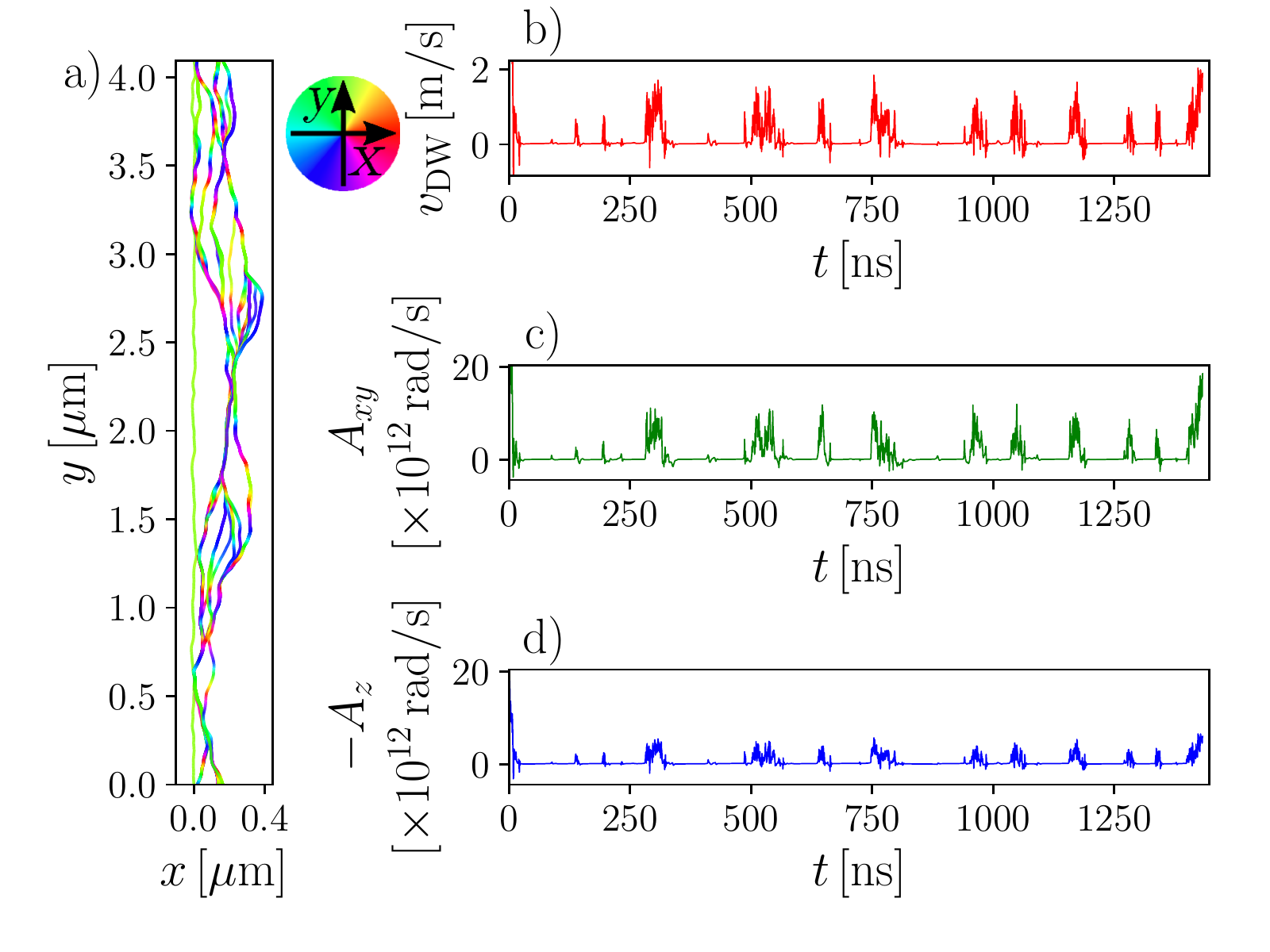}
\caption{a) An example of a sequence of DW magnetization configurations
in between successive avalanches (as defined by thresholding the $v_\mathrm{DW}$
signal); the DW is moving to the $+x$ direction. The corresponding 
crackling noise signals, with b) the DW velocity
$v_\mathrm{DW}(t)$, c) the in-plane activity $A_{xy}(t)$, and d) the out-of-plane
activity $-A_z(t)$.}
\label{fig:1}
\end{figure}

\begin{figure}[b!]
\includegraphics[trim={0 0.7cm 0 0.7cm},clip,width=.9\columnwidth]
{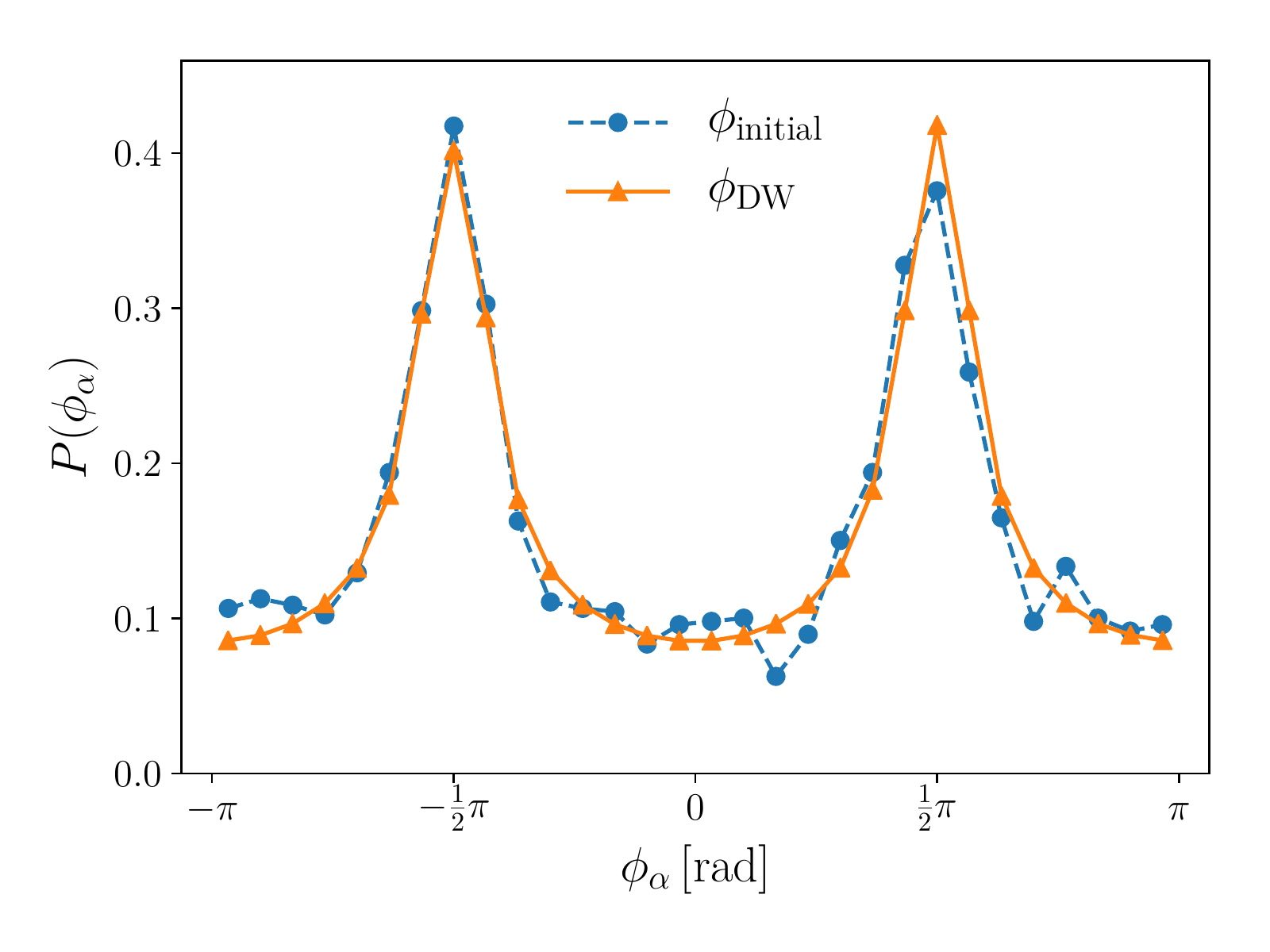}
\caption{Distribution of the in-plane magnetization angle
$\phi_\mathrm{initial}$ of the DW segments where avalanches are
initiated, vs the corresponding distribution of $\phi_\mathrm{DW}$ for 
all DW segments. The two distributions look almost identical, 
suggesting that the presence or absence of BLs within the DW is 
not important for the avalanche triggering process.}
\label{fig:init_angle}
\end{figure}

For thin films with thicknesses of only a few atoms, a natural source of 
disorder \cite{leliaert2014numerical} is given by thickness fluctuations 
of the film. Thus, for simulations of disordered films, 
the sample is divided into ``grains'' of linear size 20 nm (defining the 
disorder correlation length) by Voronoi tessellation, each grain
having a normally distributed random thickness $t_\mathrm{G} = h + 
\mathcal{N}(0,1) r h$, with $r$ 
the relative magnitude of the grain-to-grain 
thickness variations, and $h$ the mean thickness of the sample. 
These thickness fluctuations are then modeled using an approach proposed in 
Ref. \cite{moretti2017dynamical}, by modulating the saturation magnetization 
and anisotropy constant according to
$M_{\mathrm{S}}^{\mathrm{G}} = \frac{M_{\mathrm{S}} t_{\mathrm{G}}}{h}$ 
and $K_\mathrm{u}^\mathrm{G} = \frac{K_\mathrm{u} h}{t_\mathrm{G}}$.

\begin{figure*}[t!]
\centering
\includegraphics[trim={0 0.2cm 0 2.8cm},clip,width=.7\textwidth]{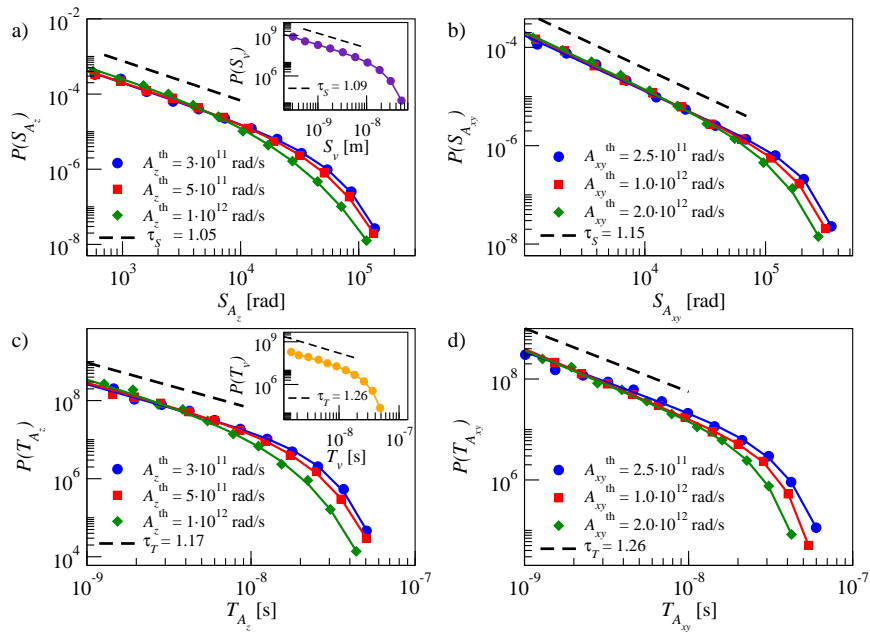}
\caption{Distributions of the avalanche sizes obtained by thresholding a) the $A_z(t)$ signal
and b) the $A_{xy}(t)$ signal. The corresponding avalanche duration distributions are shown in
c) and d), respectively. Different threshold values ($A_z^\text{th}$ and 
$A_{xy}^\text{th}$, respectively) considered are indicated in the legends. 
The insets in a) and c) show the corresponding avalanche size and duration 
distributions computed from the $v_\text{DW}(t)$ signal using 
$v_\text{DW}^\text{th}=0.1$ m/s.
Solid lines correspond to fits of power-laws terminated by a large-avalanche 
cutoff (see text), while the dashed lines show the fitted power-law exponent in each case.}
\label{fig:all}
\end{figure*}

We start by considering the response of a Bloch DW to a constant 
$B_\text{ext}$ along the $+z$ direction; this leads to DW
motion in the $+x$ direction. Our algorithm solves the spatially averaged 
DW velocity $v_\text{DW}$ by determining the local DW position along the DW as 
$X(y) = \min_y |m_z(x)|$, with $m_z(x)$ interpolated across the minimum. 
By scanning different values of the
thickness fluctuations $r$, we found that $r=0.03$ produces a similar 
$v_\mathrm{DW}(B_\mathrm{ext})$ behavior as in the finite temperature 
experiments of Ref. %by Metaxas {\it et al.} 
\cite{metaxas2007creep} for the 0.5 nm thick 
sample in the range of 0 -- 30 mT. Due to thermal rounding of
the depinning transition \cite{bustingorry2007thermal} in experiments of 
Ref. \cite{metaxas2007creep} this value of $r$ should be interpreted as
a lower limit. The resulting
$v_\mathrm{DW}(B_\mathrm{ext})$ curve is shown in Fig. \ref{fig:scan},
along with the corresponding curve from the disorder-free system. 
The depinning field of roughly $15 \: \mathrm{mT}$ due to the quenched 
pinning field exceeds the Walker threshold of $2.5 \: \mathrm{mT}$ of 
the non-disordered system, thus suggesting that the experiment of 
Ref. %Metaxas {\it et al.} 
\cite{metaxas2007creep} is operating in the precessional 
regime.

We then proceed to address the main problem of this paper, i.e., 
how Barkhausen noise is affected by the presence of BLs.
%study Barkhausen noise in the system with $r=0.03$.
%which was above shown to exhibit a depinning field well above the Walker
%field. 
%To this end, 
To this end, we consider the system with $r=0.03$, and a simulation protocol 
involving a moving
simulation window where the DW center of mass is always kept within 
one discretization cell from the center of the simulation window, using
the ext\_centerWall function of MuMax3 with a modified tolerance.
This minimizes effects due to demagnetizing fields that may slow
down the DW during avalanches. To ``re-introduce'' this feature in a 
controllable fashion, we utilize a driving protocol analogous
to the quasistatic limit of the constant velocity drive, where the driving
field $B_\text{ext}$ is decreased during avalanches (i.e., when 
$v_\text{DW}>v_\text{DW}^\text{th}=0.1$ m/s) %at a rate proportional to 
%$v_\text{DW}$, i.e., 
as $\dot{B}_\text{ext} = -k |v_\mathrm{DW}|$, with 
$k =  0.18 \, \mathrm{mT / nm}$
chosen to adjust the avalanche cutoff to be such that 
the lateral extent of the largest avalanches is smaller than $L_y$,
%the length of 
%the DW, 
in order to avoid finite size effects. In between avalanches 
(i.e., when $v_\text{DW} < 0.1$ m/s), $B_\text{ext}$ is ramped up 
at a rate $\dot{B}_\mathrm{ext} = 0.037 \, \mathrm{mT/ns}$
until the next avalanche is triggered.
The latter rate is chosen to get well-separated avalanches in time, while
at the same time avoiding excessively long waiting times between avalanches.
This leads to a $B_\text{ext}(t)$ which after an initial transient fluctuates 
in the vicinity of the depinning field.

%We modified the moving simulation window implementation of the original MuMax3 such that the center of mass of the domain wall is retained within one simulation cell from the simulation window center to minimize effects of the demagnetizing field.

%We set the initial applied field to 15 mT, which is around the depinning field of our sample. We are  driving the DW with quasistatic field analogous to pulling a weight with spring on a rough surface. We limit the velocity to 0.1 m/s. If the velocity is above the limit, the field is decreased at a rate of $k |v_\mathrm{DW}|$, where $k = 5 \times 10^{-5} \, \mathrm{s \, T / m}$ CHECK THESE AGAIN. If $v_\mathrm{DW}$ below the limit, the field is increased $1 \times 10^{-5} \, \mathrm{T/1000 timesteps}$. The algorithm solves the velocity of domain wall by determining the DW position as $X(y) = \min_y |m_z(x)|$. To get more accurate position we interpolate across the minimum. 
%The spring constant $k$ and the increase factor were balanced so that we get distinguishable avalanches from velocity signal and the wait time between avalanches is not unreasonably long.

%Given that the driving protocol described above is expected to lead to precessional
%DW dynamics during the Barkhausen jumps, we characterize the bursty DW dynamics
%by considering a number of signals. 
To characterize the bursty DW dynamics, 
in addition to the ``standard'' DW velocity $v_\text{DW}$, 
we study different measures of the rate of spin rotation (or ``activity'') 
associated with the DW dynamics. To study the dynamics of the internal 
degrees of freedom of the DW, we consider separately contributions 
from in-plane and out-of-plane spin rotation, defined as
%To this end, we analyze the in-plane and out-of-plane magnetization 
%activity signals, $A_{xy}(t)$ and $A_{z}(t)$, respectively, defined as  
%\begin{align*}
$A_{xy}(t) = \sum_{i \in B} \dot{\phi}_i \cdot |\mathbf{m}_{i,xy}|$ and
$A_{z}(t) = \sum_{i \in B} \dot{\theta}_i$, respectively,
%\end{align*}
where $\phi_i$ and $\theta_i$ are the spherical coordinate angles of the 
magnetization vector $\mathbf{m}_i$ in the $i$th discretization cell. The 
sums are taken over a band $B$ extending 20 
discretization cells around the DW on both sides, moving  
with the DW. The multiplication by $|\mathbf{m}_{i,xy}|$ in $A_{xy}$ is 
included to consider only contributions originating from inside of 
the DW. 
%The time derivatives of $\phi_i$ and $\theta_i$ are 
%calculated using finite differences on the time scale of 1000 time steps, 
%i.e., around $0.27 \: \mathrm{ns}$.

%The second term of normalization in $A_{xy}$ is used to suppress the effect in-plane components in the domains dominating in-plane activity. 

%The implementation of activity analysis and quasistatic driving forced us to run the simulation in predefined amount of time steps. We used 1000 time steps, which is around $0.5 \: \mathrm{ns}$, with Dormand-Prince solver (RK45). The time step size is adaptive. 

%\section{Results}

%We analyzed the $v_\mathrm{DW}$ under static external field $B_\mathrm{ext}$. The results are presented in Fig. \ref{fig:scan}. The simulations were done with perfect Co-film and disordered. We can see that the depinning field of $15 \: \mathrm{mT}$ is higher than the Walker field, which is around $2.5 \: \mathrm{mT}$. 

\begin{figure*}[ht!]
\centering
\includegraphics[trim={0 0.73cm 0 0.6cm},clip,width=.8\textwidth]{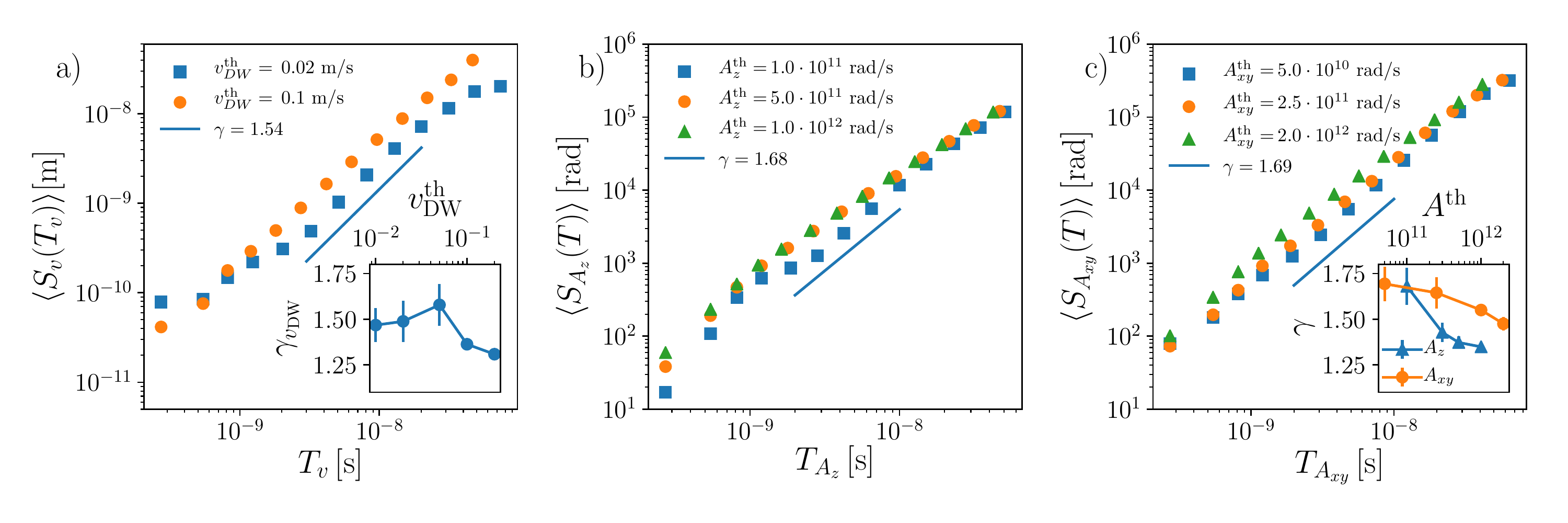}
\caption{Scaling of the average avalanche size as a function of duration for different 
threshold values. a) $\langle S_{v}(T) \rangle$, b) $\langle S_{A_z}(T) \rangle$ and 
c) $\langle S_{A_{xy}}(T) \rangle$. The insets in a) and c) illustrate the 
threshold-dependent nature of the exponent $\gamma$ characterizing the size vs duration 
scaling.}
\label{fig:TS}
\end{figure*}

Fig. \ref{fig:1}a) shows examples of DW magnetization configurations in 
between successive avalanches, defined by thresholding the $v_\text{DW}(t)$
signal with $v_\text{DW}^\text{th} = 0.1$ m/s. To quickly reach the
stationary avalanche regime, we use 15 mT as the initial field. 
Notice how the initially straight Bloch DW (green) is quickly
transformed into a rough interface with a large number of BLs, visible
in Fig. \ref{fig:1}a) as abrupt changes of color along the DW; see also Movie 1
(Supplementary Material \cite{SM}). 

Figs. \ref{fig:1}b), c) and d) show the corresponding $v_\text{DW}(t)$, 
$A_{xy}(t)$ and $-A_z(t)$ signals, respectively; notice that $A_z(t)$ has 
a minus sign to compensate for the fact that $B_\text{ext}$ along +$z$ tends 
to decrease $\theta_i$. In addition to the fact
that all three signals exhibit the characteristic bursty appearance of a 
crackling noise signal, we observe two main points: (i) $v_\text{DW}$, as well
as the two activity signals $A_{xy}(t)$ and $-A_z(t)$, may momentarily have
negative values; this indicates that the DW center of mass is moving
against the direction imposed by $B_\text{ext}$, and hence the
DW does not respect the Middleton theorem 
%applicable for elastic interfaces in random medium 
\cite{middleton1992asymptotic}. (ii) While the appearance of the three 
signals is quite similar, 
%in terms of appearence of the bursts, 
$A_{xy}(t)$ has a significantly larger magnitude than 
%the corresponding out-of-plane 
%activity 
$A_{z}(t)$: %Averaging over the activity bursts within 60 signals 
%similar to the ones shown in Fig. \ref{fig:1}
We find $\langle A_{xy} / A_z \rangle \approx 1.7$, showing that in relative
terms the BL activity within the DW is more 
pronounced during avalanches than the overall propagation of the DW. 
Comparing the distribution $P(\phi_\text{initial})$ of the local
in-plane magnetization angle $\phi_\text{initial}$ of the DW segments from which an
avalanche is triggered to that of the angle $\phi_\text{DW}$ of all DW segments
(Fig. \ref{fig:init_angle}) suggests that the avalanche triggering process is not
affected by the local DW structure.
 
%we can see that during the avalanches the magnitude of in-plane activity is higher than out-of-plane activity. It is actually higher by $\langle A_{xy} / A_z \rangle \approx 1.7$ (std. 0.5), which is averaged over 60 signals similar to the ones presented in Fig. \ref{fig:1}. Between the avalanches magnitude of $A_z$ is lower than the $A_{xy}$.

%Authors Metaxas et al. \citep{metaxas2007} were uncertain if their measured velocity was above the Walker limit. The results in Fig. \ref{fig:scan} confirm that applied field is well above the Walker limit. The figure represents average DW velocity $v_\mathrm{DW}$ as a function of static applied field. 

To analyze the statistical properties of the Barkhausen 
avalanches, we consider 200 realizations of the three signals discussed above. 
Denoting the signal by $V(t)$, the avalanche size is defined as 
$S_V = \int_0^T [V(t)-V^\text{th}]\text{d}t$,
where $V^\text{th}$ is the threshold level used to define the avalanches; the integral
is over a time interval $T$ (the avalanche duration) during which the signal stays 
continuously above $V^\text{th}$. We consider separately the three 
cases where $V(t)$
is $v_\text{DW}(t)$, $A_{z}(t)$ or $A_{xy}(t)$. Figs. \ref{fig:all} a) and b) show
the distributions $P(S_{A_z})$ and $P(S_{A_{xy}})$ for different threshold values 
($A_z^\text{th}$ and $A_{xy}^\text{th}$, respectively); The corresponding avalanche
duration distributions $P(T_{A_z})$ and $P(T_{A_{xy}})$ are shown in Figs. \ref{fig:all} 
c) and d), respectively. Insets of Figs. \ref{fig:all} a) and c) show the distributions
$P(S_v)$ and $P(T_v)$ extracted from the $v_\text{DW}$ signal using 
$v_\text{DW}^\text{th} = 0.1$ m/s. 

All the distributions can be well-described by a power law 
terminated by a large-avalanche cutoff. Solid lines in Fig \ref{fig:all} show fits of
$P(S_V) = S_V^{-\tau_{S}}\exp [-(S_V/S_V^*)^{\beta}]$, where $\tau_S$ is a scaling exponent, 
$\beta$ parametrizes the shape of the cutoff, and $S_V^*$ a cutoff avalanche size
(avalanche durations follow a similar scaling form). We find $\tau_S = 1.1 \pm 0.1$ 
and $\tau_T = 1.2 \pm 0.1$, respectively, i.e. close to the values expected for the 
quenched Edwards Wilkinson (qEW) equation, $\partial h(x,t)/\partial t = \nu \nabla^2 h(x,t) + 
\eta(x,h) + F_\text{ext}$, describing a short-range elastic string $h(x,t) $
driven by an external force $F_\text{ext}$ in a quenched random medium $\eta$ 
\cite{rosso2009avalanche}. The value of the $\tau_S$ exponent is also close to 
that found very recently for ``creep avalanches'' \cite{grassi2018intermittent}, 
and to that describing avalanches in the central hysteresis loop in a 2D RFIM 
with a built-in DW \cite{tadic2018dynamical}.
%Two main observations then follow:
%(i) The scaling exponents of the avalanche size and duration appear to be consistent
%with $\tau_S \approx 1.1$ and $\tau_T \approx 1.2$, respectively, i.e. close to the
%values expected for the quenched Edwards Wilkinson (qEW) equation describing a short-range
%elastic string in a random medium \cite{rosso2009avalanche}. (ii) 
The cutoff 
avalanche size and duration depend on the imposed threshold level, but appear to 
saturate to a value set by the ``demagnetizing factor'' $k$ in the limit of a 
low threshold. 
%Insets of Figs. \ref{fig:all} a) and c) show the distributions
%$P(S_v)$ and $P(T_v)$ extracted from the $v_\text{DW}$ signal using 0.1 m/s as the
%velocity threshold; the scaling exponents appear to be very close to the ones found
%above for the two activity signals. 
Fig. \ref{fig:TS} shows the scaling of the
average avalanche size as a function of duration, $\langle S_{v}(T)\rangle$ in a), 
$\langle S_{A_z}(T)\rangle$ in b) and $\langle S_{A_{xy}}(T)\rangle$ in c). The 
exponent $\gamma$ describing
the scaling as $\langle S_{v}(T)\rangle \sim T^{\gamma}$ (and similarly for 
$\langle S_{A_{z}}(T)\rangle$ and $\langle S_{A_{xy}}(T)\rangle$) is found to be 
threshold-dependent, in analogy to
recent observations for propagating crack lines \cite{janicevic2016interevent} 
and the RFIM \cite{janicevic2018threshold}, with the $\gamma$-value 
close to 1.6 expected for the qEW equation in the limit of zero 
threshold \cite{laurson2013evolution} approximately recovered for low thresholds 
[insets of Fig. \ref{fig:TS}a) and c)]. Thus, our exponent values satisfy within
error bars the scaling relation $\gamma = (\tau_T - 1)/(\tau_S - 1)$.
%Inset of Fig. \ref{fig:TS}a) shows the 
%$\langle S_{v}(T)\rangle$
%data extracted from the $v_\text{DW}(t)$ signal, described by a slightly smaller
%$\gamma$-exponent than that expected from the qEW equation, possibly due to the 
%finite $v_\text{DW}^\text{th} = 0.1$ m/s.

Hence, we have shown how DWs with a dynamical internal structure consisting of BLs 
generate Barkhausen noise in disordered thin films with PMA. One of the unique
features of this system is 
the large relative magnitude of the internal, in-plane bursty spin rotation activity 
within the DW, which in our case actually exceeds that of the out-of-plane spin 
rotations contributing to DW displacement. We have demonstrated
that this internal dynamics within the DW leads to a violation of the Middleton
``no-passing'' theorem. %, given that the DW is sometimes observed to move backwards.

It is quite remarkable that %we also observed that the
the scaling exponents describing the Barkhausen jumps cannot be distinguished 
from those expected for the much simpler qEW equation. %, describing an elastic 
%interface without any internal structure. 
The avalanche triggerings 
appear not to be correlated with the internal structure of the DW. 
%These findings suggest that 
Thus, commonly used simple models based on describing 
DWs as elastic interfaces, neglecting Bloch line dynamics by construction, 
seem to be capturing
correctly the large-scale critical dynamics of the system.
%lines within $1d$ DWs in $2d$ films may be irrelevant for the large-scale
%critical dynamics of the system. 
This may be rationalized by noticing that Bloch lines,
being localized N\'eel wall -like segments within the Bloch DW, produce dipolar stray 
fields decaying as $1/r^3$ in real space. For $1d$ interfaces, 
such interactions are short-ranged, and hence are not expected to change the
universality class of the avalanche dynamics from that of systems with purely 
local elasticity. In higher dimensions dipolar interactions are long-ranged,
so we expect that the internal dynamics of the DWs will have important 
consequences; the role of Bloch lines in the case of $3d$ magnets with $2d$ DWs 
should be addressed in future studies.
%This may change in higher dimensions, as dipolar interactions are 
%long-ranged for $2d$ interfaces in $3d$ random media; the role of 
%Bloch lines in such systems should be addressed in future studies.
Another important future avenue of research of great current interest would be to
%, ranging from addressing
%the role of Bloch lines in $2d$ DWs in $3d$ magnets on the Barkhausen effect to 
extend the present study to thin films with Dzyaloshinskii-Moriya interactions 
\cite{thiaville2012dynamics,yoshimura2016soliton}. 

\begin{acknowledgments}
This work has been supported by the Academy of Finland through 
an Academy Research Fellowship (LL, project no. 268302). We acknowledge the 
computational resources provided by the Aalto University School of Science 
``Science-IT'' project, as well as those provided by CSC (Finland).
\end{acknowledgments}

\bibliography{bibl}

\begin{thebibliography}{35}%
\makeatletter
\providecommand \@ifxundefined [1]{%
 \@ifx{#1\undefined}
}%
\providecommand \@ifnum [1]{%
 \ifnum #1\expandafter \@firstoftwo
 \else \expandafter \@secondoftwo
 \fi
}%
\providecommand \@ifx [1]{%
 \ifx #1\expandafter \@firstoftwo
 \else \expandafter \@secondoftwo
 \fi
}%
\providecommand \natexlab [1]{#1}%
\providecommand \enquote  [1]{``#1''}%
\providecommand \bibnamefont  [1]{#1}%
\providecommand \bibfnamefont [1]{#1}%
\providecommand \citenamefont [1]{#1}%
\providecommand \href@noop [0]{\@secondoftwo}%
\providecommand \href [0]{\begingroup \@sanitize@url \@href}%
\providecommand \@href[1]{\@@startlink{#1}\@@href}%
\providecommand \@@href[1]{\endgroup#1\@@endlink}%
\providecommand \@sanitize@url [0]{\catcode `\\12\catcode `\$12\catcode
  `\&12\catcode `\#12\catcode `\^12\catcode `\_12\catcode `\%12\relax}%
\providecommand \@@startlink[1]{}%
\providecommand \@@endlink[0]{}%
\providecommand \url  [0]{\begingroup\@sanitize@url \@url }%
\providecommand \@url [1]{\endgroup\@href {#1}{\urlprefix }}%
\providecommand \urlprefix  [0]{URL }%
\providecommand \Eprint [0]{\href }%
\providecommand \doibase [0]{http://dx.doi.org/}%
\providecommand \selectlanguage [0]{\@gobble}%
\providecommand \bibinfo  [0]{\@secondoftwo}%
\providecommand \bibfield  [0]{\@secondoftwo}%
\providecommand \translation [1]{[#1]}%
\providecommand \BibitemOpen [0]{}%
\providecommand \bibitemStop [0]{}%
\providecommand \bibitemNoStop [0]{.\EOS\space}%
\providecommand \EOS [0]{\spacefactor3000\relax}%
\providecommand \BibitemShut  [1]{\csname bibitem#1\endcsname}%
\let\auto@bib@innerbib\@empty
%</preamble>
\bibitem [{\citenamefont {Durin}\ and\ \citenamefont
  {Zapperi}(2006)}]{durin2006barkhausen}%
  \BibitemOpen
  \bibfield  {author} {\bibinfo {author} {\bibfnamefont {G.}~\bibnamefont
  {Durin}}\ and\ \bibinfo {author} {\bibfnamefont {S.}~\bibnamefont
  {Zapperi}},\ }in\ \href@noop {} {\emph {\bibinfo {booktitle} {The Science of
  Hysteresis}}},\ \bibinfo {editor} {edited by\ \bibinfo {editor}
  {\bibfnamefont {G.}~\bibnamefont {Bertotti}}\ and\ \bibinfo {editor}
  {\bibfnamefont {I.}~\bibnamefont {Mayergoyz}}}\ (\bibinfo  {publisher}
  {Academic, Amsterdam},\ \bibinfo {year} {2006})\BibitemShut {NoStop}%
\bibitem [{\citenamefont {Laurson}\ \emph {et~al.}(2013)\citenamefont
  {Laurson}, \citenamefont {Illa}, \citenamefont {Santucci}, \citenamefont
  {Tallakstad}, \citenamefont {M{\aa}l{\o}y},\ and\ \citenamefont
  {Alava}}]{laurson2013evolution}%
  \BibitemOpen
  \bibfield  {author} {\bibinfo {author} {\bibfnamefont {L.}~\bibnamefont
  {Laurson}}, \bibinfo {author} {\bibfnamefont {X.}~\bibnamefont {Illa}},
  \bibinfo {author} {\bibfnamefont {S.}~\bibnamefont {Santucci}}, \bibinfo
  {author} {\bibfnamefont {K.~T.}\ \bibnamefont {Tallakstad}}, \bibinfo
  {author} {\bibfnamefont {K.~J.}\ \bibnamefont {M{\aa}l{\o}y}}, \ and\
  \bibinfo {author} {\bibfnamefont {M.~J.}\ \bibnamefont {Alava}},\ }\href@noop
  {} {\bibfield  {journal} {\bibinfo  {journal} {Nat. commun.}\ }\textbf
  {\bibinfo {volume} {4}},\ \bibinfo {pages} {2927} (\bibinfo {year}
  {2013})}\BibitemShut {NoStop}%
\bibitem [{\citenamefont {Rost}\ \emph {et~al.}(2007)\citenamefont {Rost},
  \citenamefont {Laurson}, \citenamefont {Dub{\'e}},\ and\ \citenamefont
  {Alava}}]{rost2007fluctuations}%
  \BibitemOpen
  \bibfield  {author} {\bibinfo {author} {\bibfnamefont {M.}~\bibnamefont
  {Rost}}, \bibinfo {author} {\bibfnamefont {L.}~\bibnamefont {Laurson}},
  \bibinfo {author} {\bibfnamefont {M.}~\bibnamefont {Dub{\'e}}}, \ and\
  \bibinfo {author} {\bibfnamefont {M.}~\bibnamefont {Alava}},\ }\href@noop {}
  {\bibfield  {journal} {\bibinfo  {journal} {Phys. Rev. Lett.}\ }\textbf
  {\bibinfo {volume} {98}},\ \bibinfo {pages} {054502} (\bibinfo {year}
  {2007})}\BibitemShut {NoStop}%
\bibitem [{\citenamefont {Zapperi}\ \emph {et~al.}(1998)\citenamefont
  {Zapperi}, \citenamefont {Cizeau}, \citenamefont {Durin},\ and\ \citenamefont
  {Stanley}}]{zapperi1998dynamics}%
  \BibitemOpen
  \bibfield  {author} {\bibinfo {author} {\bibfnamefont {S.}~\bibnamefont
  {Zapperi}}, \bibinfo {author} {\bibfnamefont {P.}~\bibnamefont {Cizeau}},
  \bibinfo {author} {\bibfnamefont {G.}~\bibnamefont {Durin}}, \ and\ \bibinfo
  {author} {\bibfnamefont {H.~E.}\ \bibnamefont {Stanley}},\ }\href@noop {}
  {\bibfield  {journal} {\bibinfo  {journal} {Phys. Rev. B}\ }\textbf {\bibinfo
  {volume} {58}},\ \bibinfo {pages} {6353} (\bibinfo {year}
  {1998})}\BibitemShut {NoStop}%
\bibitem [{\citenamefont {Barkhausen}(1919)}]{barkhausen1919zwei}%
  \BibitemOpen
  \bibfield  {author} {\bibinfo {author} {\bibfnamefont {H.}~\bibnamefont
  {Barkhausen}},\ }\href@noop {} {\bibfield  {journal} {\bibinfo  {journal}
  {Phys. Z}\ }\textbf {\bibinfo {volume} {20}},\ \bibinfo {pages} {401}
  (\bibinfo {year} {1919})}\BibitemShut {NoStop}%
\bibitem [{\citenamefont {Lecomte}\ \emph {et~al.}(2009)\citenamefont
  {Lecomte}, \citenamefont {Barnes}, \citenamefont {Eckmann},\ and\
  \citenamefont {Giamarchi}}]{lecomte2009depinning}%
  \BibitemOpen
  \bibfield  {author} {\bibinfo {author} {\bibfnamefont {V.}~\bibnamefont
  {Lecomte}}, \bibinfo {author} {\bibfnamefont {S.~E.}\ \bibnamefont {Barnes}},
  \bibinfo {author} {\bibfnamefont {J.-P.}\ \bibnamefont {Eckmann}}, \ and\
  \bibinfo {author} {\bibfnamefont {T.}~\bibnamefont {Giamarchi}},\ }\href@noop
  {} {\bibfield  {journal} {\bibinfo  {journal} {Phys. Rev. B}\ }\textbf
  {\bibinfo {volume} {80}},\ \bibinfo {pages} {054413} (\bibinfo {year}
  {2009})}\BibitemShut {NoStop}%
\bibitem [{\citenamefont {Parkin}\ \emph {et~al.}(2008)\citenamefont {Parkin},
  \citenamefont {Hayashi},\ and\ \citenamefont {Thomas}}]{parkin2008magnetic}%
  \BibitemOpen
  \bibfield  {author} {\bibinfo {author} {\bibfnamefont {S.~S.}\ \bibnamefont
  {Parkin}}, \bibinfo {author} {\bibfnamefont {M.}~\bibnamefont {Hayashi}}, \
  and\ \bibinfo {author} {\bibfnamefont {L.}~\bibnamefont {Thomas}},\
  }\href@noop {} {\bibfield  {journal} {\bibinfo  {journal} {Science}\ }\textbf
  {\bibinfo {volume} {320}},\ \bibinfo {pages} {190} (\bibinfo {year}
  {2008})}\BibitemShut {NoStop}%
\bibitem [{\citenamefont {Schryer}\ and\ \citenamefont
  {Walker}(1974)}]{schryer1974motion}%
  \BibitemOpen
  \bibfield  {author} {\bibinfo {author} {\bibfnamefont {N.~L.}\ \bibnamefont
  {Schryer}}\ and\ \bibinfo {author} {\bibfnamefont {L.~R.}\ \bibnamefont
  {Walker}},\ }\href@noop {} {\bibfield  {journal} {\bibinfo  {journal} {J.
  Appl. Phys.}\ }\textbf {\bibinfo {volume} {45}},\ \bibinfo {pages} {5406}
  (\bibinfo {year} {1974})}\BibitemShut {NoStop}%
\bibitem [{\citenamefont {Mougin}\ \emph {et~al.}(2007)\citenamefont {Mougin},
  \citenamefont {Cormier}, \citenamefont {Adam}, \citenamefont {Metaxas},\ and\
  \citenamefont {Ferr{\'e}}}]{mougin2007domain}%
  \BibitemOpen
  \bibfield  {author} {\bibinfo {author} {\bibfnamefont {A.}~\bibnamefont
  {Mougin}}, \bibinfo {author} {\bibfnamefont {M.}~\bibnamefont {Cormier}},
  \bibinfo {author} {\bibfnamefont {J.}~\bibnamefont {Adam}}, \bibinfo {author}
  {\bibfnamefont {P.}~\bibnamefont {Metaxas}}, \ and\ \bibinfo {author}
  {\bibfnamefont {J.}~\bibnamefont {Ferr{\'e}}},\ }\href@noop {} {\bibfield
  {journal} {\bibinfo  {journal} {EPL}\ }\textbf {\bibinfo {volume} {78}},\
  \bibinfo {pages} {57007} (\bibinfo {year} {2007})}\BibitemShut {NoStop}%
\bibitem [{\citenamefont {Thiaville}\ and\ \citenamefont
  {Nakatani}(2006)}]{thiaville2006domain}%
  \BibitemOpen
  \bibfield  {author} {\bibinfo {author} {\bibfnamefont {A.}~\bibnamefont
  {Thiaville}}\ and\ \bibinfo {author} {\bibfnamefont {Y.}~\bibnamefont
  {Nakatani}},\ }in\ \href@noop {} {\emph {\bibinfo {booktitle} {Spin dynamics
  in confined magnetic structures III}}}\ (\bibinfo  {publisher} {Springer},\
  \bibinfo {year} {2006})\ pp.\ \bibinfo {pages} {161--205}\BibitemShut
  {NoStop}%
\bibitem [{\citenamefont {Herranen}\ and\ \citenamefont
  {Laurson}(2015)}]{herranen2015domain}%
  \BibitemOpen
  \bibfield  {author} {\bibinfo {author} {\bibfnamefont {T.}~\bibnamefont
  {Herranen}}\ and\ \bibinfo {author} {\bibfnamefont {L.}~\bibnamefont
  {Laurson}},\ }\href@noop {} {\bibfield  {journal} {\bibinfo  {journal} {Phys.
  Rev. B}\ }\textbf {\bibinfo {volume} {92}},\ \bibinfo {pages} {100405}
  (\bibinfo {year} {2015})}\BibitemShut {NoStop}%
\bibitem [{\citenamefont {Herranen}\ and\ \citenamefont
  {Laurson}(2017)}]{herranen2017bloch}%
  \BibitemOpen
  \bibfield  {author} {\bibinfo {author} {\bibfnamefont {T.}~\bibnamefont
  {Herranen}}\ and\ \bibinfo {author} {\bibfnamefont {L.}~\bibnamefont
  {Laurson}},\ }\href@noop {} {\bibfield  {journal} {\bibinfo  {journal} {Phys.
  Rev. B}\ }\textbf {\bibinfo {volume} {96}},\ \bibinfo {pages} {144422}
  (\bibinfo {year} {2017})}\BibitemShut {NoStop}%
\bibitem [{\citenamefont {Malozemoff}\ and\ \citenamefont
  {Slonczewski}(1979)}]{malozemoff1979magnetic}%
  \BibitemOpen
  \bibfield  {author} {\bibinfo {author} {\bibfnamefont {A.}~\bibnamefont
  {Malozemoff}}\ and\ \bibinfo {author} {\bibfnamefont {J.}~\bibnamefont
  {Slonczewski}},\ }\href@noop {} {\emph {\bibinfo {title} {Magnetic Domain
  Walls in Bubble Materials}}}\ (\bibinfo  {publisher} {Academic press},\
  \bibinfo {year} {1979})\BibitemShut {NoStop}%
\bibitem [{\citenamefont {Laurson}\ \emph {et~al.}(2014)\citenamefont
  {Laurson}, \citenamefont {Durin},\ and\ \citenamefont
  {Zapperi}}]{laurson2014universality}%
  \BibitemOpen
  \bibfield  {author} {\bibinfo {author} {\bibfnamefont {L.}~\bibnamefont
  {Laurson}}, \bibinfo {author} {\bibfnamefont {G.}~\bibnamefont {Durin}}, \
  and\ \bibinfo {author} {\bibfnamefont {S.}~\bibnamefont {Zapperi}},\
  }\href@noop {} {\bibfield  {journal} {\bibinfo  {journal} {Phys. Rev. B}\
  }\textbf {\bibinfo {volume} {89}},\ \bibinfo {pages} {104402} (\bibinfo
  {year} {2014})}\BibitemShut {NoStop}%
\bibitem [{\citenamefont {Caballero}\ \emph {et~al.}(2018)\citenamefont
  {Caballero}, \citenamefont {Ferrero}, \citenamefont {Kolton}, \citenamefont
  {Curiale}, \citenamefont {Jeudy},\ and\ \citenamefont
  {Bustingorry}}]{caballero2018magnetic}%
  \BibitemOpen
  \bibfield  {author} {\bibinfo {author} {\bibfnamefont {N.~B.}\ \bibnamefont
  {Caballero}}, \bibinfo {author} {\bibfnamefont {E.~E.}\ \bibnamefont
  {Ferrero}}, \bibinfo {author} {\bibfnamefont {A.~B.}\ \bibnamefont {Kolton}},
  \bibinfo {author} {\bibfnamefont {J.}~\bibnamefont {Curiale}}, \bibinfo
  {author} {\bibfnamefont {V.}~\bibnamefont {Jeudy}}, \ and\ \bibinfo {author}
  {\bibfnamefont {S.}~\bibnamefont {Bustingorry}},\ }\href@noop {} {\bibfield
  {journal} {\bibinfo  {journal} {Physical Review E}\ }\textbf {\bibinfo
  {volume} {97}},\ \bibinfo {pages} {062122} (\bibinfo {year}
  {2018})}\BibitemShut {NoStop}%
\bibitem [{\citenamefont {P{\'e}rez-Reche}\ and\ \citenamefont
  {Vives}(2004)}]{perez2004spanning}%
  \BibitemOpen
  \bibfield  {author} {\bibinfo {author} {\bibfnamefont {F.~J.}\ \bibnamefont
  {P{\'e}rez-Reche}}\ and\ \bibinfo {author} {\bibfnamefont {E.}~\bibnamefont
  {Vives}},\ }\href@noop {} {\bibfield  {journal} {\bibinfo  {journal}
  {Physical Review B}\ }\textbf {\bibinfo {volume} {70}},\ \bibinfo {pages}
  {214422} (\bibinfo {year} {2004})}\BibitemShut {NoStop}%
\bibitem [{\citenamefont {Mughal}\ \emph {et~al.}(2010)\citenamefont {Mughal},
  \citenamefont {Laurson}, \citenamefont {Durin},\ and\ \citenamefont
  {Zapperi}}]{mughal2010effect}%
  \BibitemOpen
  \bibfield  {author} {\bibinfo {author} {\bibfnamefont {A.}~\bibnamefont
  {Mughal}}, \bibinfo {author} {\bibfnamefont {L.}~\bibnamefont {Laurson}},
  \bibinfo {author} {\bibfnamefont {G.}~\bibnamefont {Durin}}, \ and\ \bibinfo
  {author} {\bibfnamefont {S.}~\bibnamefont {Zapperi}},\ }\href@noop {}
  {\bibfield  {journal} {\bibinfo  {journal} {IEEE Trans. Magn.}\ }\textbf
  {\bibinfo {volume} {46}},\ \bibinfo {pages} {228} (\bibinfo {year}
  {2010})}\BibitemShut {NoStop}%
\bibitem [{\citenamefont {Spasojevi{\'c}}\ \emph {et~al.}(2011)\citenamefont
  {Spasojevi{\'c}}, \citenamefont {Jani{\'c}evi{\'c}},\ and\ \citenamefont
  {Kne{\v{z}}evi{\'c}}}]{spasojevic2011avalanche}%
  \BibitemOpen
  \bibfield  {author} {\bibinfo {author} {\bibfnamefont {D.}~\bibnamefont
  {Spasojevi{\'c}}}, \bibinfo {author} {\bibfnamefont {S.}~\bibnamefont
  {Jani{\'c}evi{\'c}}}, \ and\ \bibinfo {author} {\bibfnamefont
  {M.}~\bibnamefont {Kne{\v{z}}evi{\'c}}},\ }\href@noop {} {\bibfield
  {journal} {\bibinfo  {journal} {Physical Review E}\ }\textbf {\bibinfo
  {volume} {84}},\ \bibinfo {pages} {051119} (\bibinfo {year}
  {2011})}\BibitemShut {NoStop}%
\bibitem [{\citenamefont {Metaxas}\ \emph {et~al.}(2007)\citenamefont
  {Metaxas}, \citenamefont {Jamet}, \citenamefont {Mougin}, \citenamefont
  {Cormier}, \citenamefont {Ferr{\'e}}, \citenamefont {Baltz}, \citenamefont
  {Rodmacq}, \citenamefont {Dieny},\ and\ \citenamefont
  {Stamps}}]{metaxas2007creep}%
  \BibitemOpen
  \bibfield  {author} {\bibinfo {author} {\bibfnamefont {P.}~\bibnamefont
  {Metaxas}}, \bibinfo {author} {\bibfnamefont {J.}~\bibnamefont {Jamet}},
  \bibinfo {author} {\bibfnamefont {A.}~\bibnamefont {Mougin}}, \bibinfo
  {author} {\bibfnamefont {M.}~\bibnamefont {Cormier}}, \bibinfo {author}
  {\bibfnamefont {J.}~\bibnamefont {Ferr{\'e}}}, \bibinfo {author}
  {\bibfnamefont {V.}~\bibnamefont {Baltz}}, \bibinfo {author} {\bibfnamefont
  {B.}~\bibnamefont {Rodmacq}}, \bibinfo {author} {\bibfnamefont
  {B.}~\bibnamefont {Dieny}}, \ and\ \bibinfo {author} {\bibfnamefont
  {R.}~\bibnamefont {Stamps}},\ }\href@noop {} {\bibfield  {journal} {\bibinfo
  {journal} {Phys. Rev. Lett.}\ }\textbf {\bibinfo {volume} {99}},\ \bibinfo
  {pages} {217208} (\bibinfo {year} {2007})}\BibitemShut {NoStop}%
\bibitem [{\citenamefont {Kim}\ \emph {et~al.}(2003)\citenamefont {Kim},
  \citenamefont {Choe},\ and\ \citenamefont {Shin}}]{kim2003direct}%
  \BibitemOpen
  \bibfield  {author} {\bibinfo {author} {\bibfnamefont {D.-H.}\ \bibnamefont
  {Kim}}, \bibinfo {author} {\bibfnamefont {S.-B.}\ \bibnamefont {Choe}}, \
  and\ \bibinfo {author} {\bibfnamefont {S.-C.}\ \bibnamefont {Shin}},\
  }\href@noop {} {\bibfield  {journal} {\bibinfo  {journal} {Phys. Rev. Lett.}\
  }\textbf {\bibinfo {volume} {90}},\ \bibinfo {pages} {087203} (\bibinfo
  {year} {2003})}\BibitemShut {NoStop}%
\bibitem [{\citenamefont {Papanikolaou}\ \emph {et~al.}(2011)\citenamefont
  {Papanikolaou}, \citenamefont {Bohn}, \citenamefont {Sommer}, \citenamefont
  {Durin}, \citenamefont {Zapperi},\ and\ \citenamefont
  {Sethna}}]{papanikolaou2011universality}%
  \BibitemOpen
  \bibfield  {author} {\bibinfo {author} {\bibfnamefont {S.}~\bibnamefont
  {Papanikolaou}}, \bibinfo {author} {\bibfnamefont {F.}~\bibnamefont {Bohn}},
  \bibinfo {author} {\bibfnamefont {R.~L.}\ \bibnamefont {Sommer}}, \bibinfo
  {author} {\bibfnamefont {G.}~\bibnamefont {Durin}}, \bibinfo {author}
  {\bibfnamefont {S.}~\bibnamefont {Zapperi}}, \ and\ \bibinfo {author}
  {\bibfnamefont {J.~P.}\ \bibnamefont {Sethna}},\ }\href@noop {} {\bibfield
  {journal} {\bibinfo  {journal} {Nat. Phys.}\ }\textbf {\bibinfo {volume}
  {7}},\ \bibinfo {pages} {316} (\bibinfo {year} {2011})}\BibitemShut {NoStop}%
\bibitem [{\citenamefont {Middleton}(1992)}]{middleton1992asymptotic}%
  \BibitemOpen
  \bibfield  {author} {\bibinfo {author} {\bibfnamefont {A.~A.}\ \bibnamefont
  {Middleton}},\ }\href@noop {} {\bibfield  {journal} {\bibinfo  {journal}
  {Phys. Rev. Lett.}\ }\textbf {\bibinfo {volume} {68}},\ \bibinfo {pages}
  {670} (\bibinfo {year} {1992})}\BibitemShut {NoStop}%
\bibitem [{\citenamefont {Le~Doussal}\ \emph {et~al.}(2002)\citenamefont
  {Le~Doussal}, \citenamefont {Wiese},\ and\ \citenamefont
  {Chauve}}]{le2002two}%
  \BibitemOpen
  \bibfield  {author} {\bibinfo {author} {\bibfnamefont {P.}~\bibnamefont
  {Le~Doussal}}, \bibinfo {author} {\bibfnamefont {K.~J.}\ \bibnamefont
  {Wiese}}, \ and\ \bibinfo {author} {\bibfnamefont {P.}~\bibnamefont
  {Chauve}},\ }\href@noop {} {\bibfield  {journal} {\bibinfo  {journal} {Phys.
  Rev. B}\ }\textbf {\bibinfo {volume} {66}},\ \bibinfo {pages} {174201}
  (\bibinfo {year} {2002})}\BibitemShut {NoStop}%
\bibitem [{\citenamefont {Vansteenkiste}\ \emph {et~al.}(2014)\citenamefont
  {Vansteenkiste}, \citenamefont {Leliaert}, \citenamefont {Dvornik},
  \citenamefont {Helsen}, \citenamefont {Garcia-Sanchez},\ and\ \citenamefont
  {Van~Waeyenberge}}]{vansteenkiste2014design}%
  \BibitemOpen
  \bibfield  {author} {\bibinfo {author} {\bibfnamefont {A.}~\bibnamefont
  {Vansteenkiste}}, \bibinfo {author} {\bibfnamefont {J.}~\bibnamefont
  {Leliaert}}, \bibinfo {author} {\bibfnamefont {M.}~\bibnamefont {Dvornik}},
  \bibinfo {author} {\bibfnamefont {M.}~\bibnamefont {Helsen}}, \bibinfo
  {author} {\bibfnamefont {F.}~\bibnamefont {Garcia-Sanchez}}, \ and\ \bibinfo
  {author} {\bibfnamefont {B.}~\bibnamefont {Van~Waeyenberge}},\ }\href@noop {}
  {\bibfield  {journal} {\bibinfo  {journal} {AIP advances}\ }\textbf {\bibinfo
  {volume} {4}},\ \bibinfo {pages} {107133} (\bibinfo {year}
  {2014})}\BibitemShut {NoStop}%
\bibitem [{\citenamefont {Leliaert}\ \emph {et~al.}(2014)\citenamefont
  {Leliaert}, \citenamefont {Van~de Wiele}, \citenamefont {Vansteenkiste},
  \citenamefont {Laurson}, \citenamefont {Durin}, \citenamefont {Dupr{\'e}},\
  and\ \citenamefont {Van~Waeyenberge}}]{leliaert2014numerical}%
  \BibitemOpen
  \bibfield  {author} {\bibinfo {author} {\bibfnamefont {J.}~\bibnamefont
  {Leliaert}}, \bibinfo {author} {\bibfnamefont {B.}~\bibnamefont {Van~de
  Wiele}}, \bibinfo {author} {\bibfnamefont {A.}~\bibnamefont {Vansteenkiste}},
  \bibinfo {author} {\bibfnamefont {L.}~\bibnamefont {Laurson}}, \bibinfo
  {author} {\bibfnamefont {G.}~\bibnamefont {Durin}}, \bibinfo {author}
  {\bibfnamefont {L.}~\bibnamefont {Dupr{\'e}}}, \ and\ \bibinfo {author}
  {\bibfnamefont {B.}~\bibnamefont {Van~Waeyenberge}},\ }\href@noop {}
  {\bibfield  {journal} {\bibinfo  {journal} {Journal of Applied Physics}\
  }\textbf {\bibinfo {volume} {115}},\ \bibinfo {pages} {17D102} (\bibinfo
  {year} {2014})}\BibitemShut {NoStop}%
\bibitem [{\citenamefont {Moretti}\ \emph {et~al.}(2017)\citenamefont
  {Moretti}, \citenamefont {Voto},\ and\ \citenamefont
  {Martinez}}]{moretti2017dynamical}%
  \BibitemOpen
  \bibfield  {author} {\bibinfo {author} {\bibfnamefont {S.}~\bibnamefont
  {Moretti}}, \bibinfo {author} {\bibfnamefont {M.}~\bibnamefont {Voto}}, \
  and\ \bibinfo {author} {\bibfnamefont {E.}~\bibnamefont {Martinez}},\
  }\href@noop {} {\bibfield  {journal} {\bibinfo  {journal} {Phys. Rev. B}\
  }\textbf {\bibinfo {volume} {96}},\ \bibinfo {pages} {054433} (\bibinfo
  {year} {2017})}\BibitemShut {NoStop}%
\bibitem [{\citenamefont {Bustingorry}\ \emph {et~al.}(2007)\citenamefont
  {Bustingorry}, \citenamefont {Kolton},\ and\ \citenamefont
  {Giamarchi}}]{bustingorry2007thermal}%
  \BibitemOpen
  \bibfield  {author} {\bibinfo {author} {\bibfnamefont {S.}~\bibnamefont
  {Bustingorry}}, \bibinfo {author} {\bibfnamefont {A.}~\bibnamefont {Kolton}},
  \ and\ \bibinfo {author} {\bibfnamefont {T.}~\bibnamefont {Giamarchi}},\
  }\href@noop {} {\bibfield  {journal} {\bibinfo  {journal} {EPL}\ }\textbf
  {\bibinfo {volume} {81}},\ \bibinfo {pages} {26005} (\bibinfo {year}
  {2007})}\BibitemShut {NoStop}%
\bibitem [{SM()}]{SM}%
  \BibitemOpen
  \href@noop {} {}\bibinfo {note} {See Supplemental Material at [URL will be
  inserted by the publisher] for a movie illustrating the bursty dynamics of a
  DW containing Bloch lines.}\BibitemShut {Stop}%
\bibitem [{\citenamefont {Rosso}\ \emph {et~al.}(2009)\citenamefont {Rosso},
  \citenamefont {Le~Doussal},\ and\ \citenamefont
  {Wiese}}]{rosso2009avalanche}%
  \BibitemOpen
  \bibfield  {author} {\bibinfo {author} {\bibfnamefont {A.}~\bibnamefont
  {Rosso}}, \bibinfo {author} {\bibfnamefont {P.}~\bibnamefont {Le~Doussal}}, \
  and\ \bibinfo {author} {\bibfnamefont {K.~J.}\ \bibnamefont {Wiese}},\
  }\href@noop {} {\bibfield  {journal} {\bibinfo  {journal} {Phys. Rev. B}\
  }\textbf {\bibinfo {volume} {80}},\ \bibinfo {pages} {144204} (\bibinfo
  {year} {2009})}\BibitemShut {NoStop}%
\bibitem [{\citenamefont {Grassi}\ \emph {et~al.}(2018)\citenamefont {Grassi},
  \citenamefont {Kolton}, \citenamefont {Jeudy}, \citenamefont {Mougin},
  \citenamefont {Bustingorry},\ and\ \citenamefont
  {Curiale}}]{grassi2018intermittent}%
  \BibitemOpen
  \bibfield  {author} {\bibinfo {author} {\bibfnamefont {M.~P.}\ \bibnamefont
  {Grassi}}, \bibinfo {author} {\bibfnamefont {A.~B.}\ \bibnamefont {Kolton}},
  \bibinfo {author} {\bibfnamefont {V.}~\bibnamefont {Jeudy}}, \bibinfo
  {author} {\bibfnamefont {A.}~\bibnamefont {Mougin}}, \bibinfo {author}
  {\bibfnamefont {S.}~\bibnamefont {Bustingorry}}, \ and\ \bibinfo {author}
  {\bibfnamefont {J.}~\bibnamefont {Curiale}},\ }\href@noop {} {\bibfield
  {journal} {\bibinfo  {journal} {arXiv preprint arXiv:1804.09572}\ } (\bibinfo
  {year} {2018})}\BibitemShut {NoStop}%
\bibitem [{\citenamefont {Tadi{\'c}}(2018)}]{tadic2018dynamical}%
  \BibitemOpen
  \bibfield  {author} {\bibinfo {author} {\bibfnamefont {B.}~\bibnamefont
  {Tadi{\'c}}},\ }\href@noop {} {\bibfield  {journal} {\bibinfo  {journal}
  {Physica A: Statistical Mechanics and its Applications}\ }\textbf {\bibinfo
  {volume} {493}},\ \bibinfo {pages} {330} (\bibinfo {year}
  {2018})}\BibitemShut {NoStop}%
\bibitem [{\citenamefont {Jani{\'c}evi{\'c}}\ \emph {et~al.}(2016)\citenamefont
  {Jani{\'c}evi{\'c}}, \citenamefont {Laurson}, \citenamefont {M{\aa}l{\o}y},
  \citenamefont {Santucci},\ and\ \citenamefont
  {Alava}}]{janicevic2016interevent}%
  \BibitemOpen
  \bibfield  {author} {\bibinfo {author} {\bibfnamefont {S.}~\bibnamefont
  {Jani{\'c}evi{\'c}}}, \bibinfo {author} {\bibfnamefont {L.}~\bibnamefont
  {Laurson}}, \bibinfo {author} {\bibfnamefont {K.~J.}\ \bibnamefont
  {M{\aa}l{\o}y}}, \bibinfo {author} {\bibfnamefont {S.}~\bibnamefont
  {Santucci}}, \ and\ \bibinfo {author} {\bibfnamefont {M.~J.}\ \bibnamefont
  {Alava}},\ }\href@noop {} {\bibfield  {journal} {\bibinfo  {journal} {Phys.
  Rev. Lett.}\ }\textbf {\bibinfo {volume} {117}},\ \bibinfo {pages} {230601}
  (\bibinfo {year} {2016})}\BibitemShut {NoStop}%
\bibitem [{\citenamefont {Jani{\'c}evi{\'c}}\ \emph {et~al.}(2018)\citenamefont
  {Jani{\'c}evi{\'c}}, \citenamefont {Jovkovi{\'c}}, \citenamefont {Laurson},\
  and\ \citenamefont {Spasojevi{\'c}}}]{janicevic2018threshold}%
  \BibitemOpen
  \bibfield  {author} {\bibinfo {author} {\bibfnamefont {S.}~\bibnamefont
  {Jani{\'c}evi{\'c}}}, \bibinfo {author} {\bibfnamefont {D.}~\bibnamefont
  {Jovkovi{\'c}}}, \bibinfo {author} {\bibfnamefont {L.}~\bibnamefont
  {Laurson}}, \ and\ \bibinfo {author} {\bibfnamefont {D.}~\bibnamefont
  {Spasojevi{\'c}}},\ }\href@noop {} {\bibfield  {journal} {\bibinfo  {journal}
  {Sci. Rep.}\ }\textbf {\bibinfo {volume} {8}},\ \bibinfo {pages} {2571}
  (\bibinfo {year} {2018})}\BibitemShut {NoStop}%
\bibitem [{\citenamefont {Thiaville}\ \emph {et~al.}(2012)\citenamefont
  {Thiaville}, \citenamefont {Rohart}, \citenamefont {Ju{\'e}}, \citenamefont
  {Cros},\ and\ \citenamefont {Fert}}]{thiaville2012dynamics}%
  \BibitemOpen
  \bibfield  {author} {\bibinfo {author} {\bibfnamefont {A.}~\bibnamefont
  {Thiaville}}, \bibinfo {author} {\bibfnamefont {S.}~\bibnamefont {Rohart}},
  \bibinfo {author} {\bibfnamefont {{\'E}.}~\bibnamefont {Ju{\'e}}}, \bibinfo
  {author} {\bibfnamefont {V.}~\bibnamefont {Cros}}, \ and\ \bibinfo {author}
  {\bibfnamefont {A.}~\bibnamefont {Fert}},\ }\href@noop {} {\bibfield
  {journal} {\bibinfo  {journal} {EPL}\ }\textbf {\bibinfo {volume} {100}},\
  \bibinfo {pages} {57002} (\bibinfo {year} {2012})}\BibitemShut {NoStop}%
\bibitem [{\citenamefont {Yoshimura}\ \emph {et~al.}(2016)\citenamefont
  {Yoshimura}, \citenamefont {Kim}, \citenamefont {Taniguchi}, \citenamefont
  {Tono}, \citenamefont {Ueda}, \citenamefont {Hiramatsu}, \citenamefont
  {Moriyama}, \citenamefont {Yamada}, \citenamefont {Nakatani},\ and\
  \citenamefont {Ono}}]{yoshimura2016soliton}%
  \BibitemOpen
  \bibfield  {author} {\bibinfo {author} {\bibfnamefont {Y.}~\bibnamefont
  {Yoshimura}}, \bibinfo {author} {\bibfnamefont {K.-J.}\ \bibnamefont {Kim}},
  \bibinfo {author} {\bibfnamefont {T.}~\bibnamefont {Taniguchi}}, \bibinfo
  {author} {\bibfnamefont {T.}~\bibnamefont {Tono}}, \bibinfo {author}
  {\bibfnamefont {K.}~\bibnamefont {Ueda}}, \bibinfo {author} {\bibfnamefont
  {R.}~\bibnamefont {Hiramatsu}}, \bibinfo {author} {\bibfnamefont
  {T.}~\bibnamefont {Moriyama}}, \bibinfo {author} {\bibfnamefont
  {K.}~\bibnamefont {Yamada}}, \bibinfo {author} {\bibfnamefont
  {Y.}~\bibnamefont {Nakatani}}, \ and\ \bibinfo {author} {\bibfnamefont
  {T.}~\bibnamefont {Ono}},\ }\href@noop {} {\bibfield  {journal} {\bibinfo
  {journal} {Nat. Phys.}\ }\textbf {\bibinfo {volume} {12}},\ \bibinfo {pages}
  {157} (\bibinfo {year} {2016})}\BibitemShut {NoStop}%
\end{thebibliography}%

\end{document}